\documentclass{aomart}

\usepackage[utf8]{inputenc}
\usepackage[english]{babel}
\usepackage{booktabs}
\usepackage{amsmath}
\usepackage{mathrsfs}
\usepackage{subfigure}
\usepackage{graphicx,lipsum}
\usepackage[toc,page]{appendix}
\usepackage{hyperref}
\usepackage{enumerate}
\usepackage{tabularx}

\makeatletter
\copyrightnote{\textcopyright~2023 by the author.}

\title{A Unified Framework for Total Variation Regularized Optimization in Fluid Dynamics and Related Physical Systems}

\newcommand{\authorname}{Varsha Gupta }

\hypersetup{
  pdftitle={A Unified Framework for Total Variation Regularized Optimization in Fluid Dynamics and Related Physical Systems},
  pdfauthor={\authorname}
}

\author{\authorname{\href{mailto:vvarsha@purdue.edu}{vvarsha@purdue.edu}}}
\fulladdress{Purdue University, West Lafayette, Indiana, United States}

\newenvironment{conditions*}
  {\par\vspace{\abovedisplayskip}\noindent
   \tabularx{\columnwidth}{>{$}l<{$} @{${}={}$} >{\raggedright\arraybackslash}X}}
  {\endtabularx\par\vspace{\belowdisplayskip}}

\startpage{1}
\endpage{}

\begin{document}

\thispagestyle{empty}
\maketitle
 \begin{abstract}

 An optimization framework is presented for minimizing the energy functional developed around a generalized equation governing physical systems such as fluid dynamics, particle transport, phase transition, and other related systems. The convexity of the energy functional is investigated to derive the necessary conditions for a smooth and global optimum solution. Furthermore, the Total Variation (TV) regularization term is introduced to gain insights into the solution space and convergence analysis of convection-dominated problems. 
 We demonstrate the practical application of our method by applying it to some selected examples such as the Boltzmann, Navier-Stokes, and Maxwell equations.
\end{abstract}

\tableofcontents
\thispagestyle{plain}
\section{Introduction}
The behavior of velocity vector field in fluid dynamics \cite{lukaszewicz2016navier}, electric and magnetic vector fields in electromagnetism \cite{haus1989electromagnetic}, and the evolution of particle distribution function in statistical mechanics \cite{davidson2013statistical} are governed by a system of partial differential equations (PDEs) \cite{evans2022partial}. These physical systems and other related systems such as magnetization dynamics \cite{lakshmanan2011fascinating} and stability \& bifurcation theory \cite{sirovich1987ginzburg} are modeled by the governing equations exhibiting certain structural symmetry \cite{chemin2011global, klainerman2010pde}. The partial derivative of the field variable is related to the self-interaction or convective term, diffusion term, and system-specific force terms. 

The presence of non-linearity makes it challenging to solve these systems of equations analytically or to gain an understanding of the nature of the solutions. Specifically, in fluid mechanics, proving the existence of global optimality and smoothness of the solution of the Navier Stokes (NS) equation remains challenging \cite{doering20093d}. A significant body of theoretical advancement has been made towards proving global solutions and regularity under different conditions. For instance, the existence of a strong global and unique solution has been proved in the case of isentropic compressible fluids with initial data close to a stable equilibrium in critical spaces, particularly, in Besov space \cite{danchin2000global}. The proof builds upon homogenous Littlewood-Paley decomposition and consequently, quasi orthogonality. Subsequently, using the decay estimates for a mixed parabolic/ hyperbolic linear system of equations \cite{hoff1995multi} and smoothness of coupled heat equation led to dissecting the investigation into low and high-frequency cases and approximating the solution using an iterative process. This proof of global existence and uniqueness is further extended to the small perturbation in initial velocity in barotropic compressible flow and for large perturbation about stable equilibrium for incompressible flow in $L^p$ type Besov space \cite{charve2010global}. For non-decaying and bounded initial velocities, the existence of a unique, global in-time solution was proved using Friedrich's approximation of the NS equations and decay estimate of the Fourier transformation of the solution\cite{giga2001global}. For the three-dimensional incompressible NS equation, the proof of global smooth solution was established for the cases when the norm takes arbitrarily large values \cite{chemin2006global} and when the norm tends to blow up \cite{chemin2011global}.

In addition to Harmonic analysis \cite{cannone2004harmonic}, the models governing these physical systems have long been studied using a variety of other mathematical tools to handle PDEs such as numerical methods \cite{hawken1991review, olver2014introduction} and perturbation analysis \cite{pierson1962perturbation, lipton2021perturbation}. The most commonly used numerical methods are finite volume, finite element, and finite difference methods \cite{MATTIUSSI20001}.

With the recent developments in artificial intelligence \cite{blechschmidt2021three} and quantum computing \cite{gaitan2021finding}, tools like deep learning \cite{milan2021deep} and physics-informed neural networks (PINN) \cite{chen2021physics} have also been deployed to gain insights into the nature of the solution of these problems. 

Although significant progress has been made toward studying these physical systems, the existing literature on PDEs is skewed toward tailored solutions to specific physical systems. While these studies have significantly improved our understanding of the underlying processes, they do little to describe a broader class of problems or offer insights that can be readily generalized. Therefore, there is a need to develop a systematic and unified mathematical framework across related physical systems for deriving generalized results about the global optimality and smoothness of their solutions under different conditions. 

The present work aims to study the Navier-Stokes equation and other related physical systems by developing a systematic unified method. A generalized equation is defined along with an associated energy functional that captures the essential features of the system. An optimization framework is developed to minimize the energy functional, constrained to the general equation. The energy functional and the constraint are combined using a Lagrange multiplier to form the Lagrangian. The first variation of the resultant equation is then analyzed for convexity and global optimality. The erratic behavior due to the self-interaction convective term is discussed.\\
To reduce the noise in the system, we introduce the TV regularization term to the original energy functional and re-derive the Euler-Lagrange equation \cite{birkhoff1983numerical} with the modified energy functional. \\
To optimize the field variable, we develop a flow similar to the Ricci flow \cite{chow2023hamilton}, that takes the energy functional towards lower energy states. The field variable is then updated using an iterative process such that it minimizes the energy functional with every successive iteration.  \\
We also develop tentative bounds on the solution space and perform convergence analysis. 

In the end, to illustrate the utility of the method, we apply it to different physical systems. Although we have not conducted a detailed perturbation analysis, the reader may utilize this method or alternative approaches to get an approximate solution for each system. 

\section{Generalized Governing Equation}

Consider a physical system such as fluid dynamics or other related physical systems where the classical vector field is associated with a primary field variable, $A$. We can write the following general equation to study the dynamic behavior of this physical system:

\begin{equation}\label{eq:1}
    \frac{\partial A}{\partial t} = -C(A \cdot \nabla)A + D\nabla^2 A + F(A)
\end{equation}

where $C$ and $D$ are the convective and diffusive coefficients respectively, associated with the convective and diffusive forces arising on account of non-linear transport due to self-interaction and stabilizing forces due to dissipative forces respectively. The former leads to complex erratic behavior while the latter leads to smooth solutions. The third term, $F(A)$ represents the forces specific to the physical system under consideration.

\section{Formulation of the Optimization Problem}
In order to analyze and understand the behavior of these physical systems under different scenarios, we will transform this general equation into an optimization framework.
Let's define the following energy functional, $E(A)$ as the objective function:

\begin{equation}
E(A) = \frac{1}{2} \int_{\Omega} \left[ |A|^2 + C |A|^2 |\nabla A|^2+ D|\nabla A|^2 \right] dV - \int_{\Omega} f(A) dV
\end{equation}

Where f(A) is the anti-derivative of F w.r.t. $A$.
The objective is to find $A$ that minimizes $E(A)$ subject to the constraint equation \eqref{eq:1}.

To gain insight into the above optimization problem, let's develop the Lagrange equation. We incorporate the constraint into the objective function using a Lagrange multiplier $\lambda$. Therefore, the augmented functional is given by:
\begin{equation}
    \begin{aligned}
\mathcal{L}(A, \lambda) = E(A) + \int_{\Omega} \lambda \left(\frac{\partial A}{\partial t} + C(A \cdot \nabla)A - D\nabla^2 A - F(A)\right) dV
\end{aligned}
\end{equation}

The proposed optimization framework can provide deeper insights into the nature of solutions of Navier-Stokes equations. The convexity of the energy functional determines if there is a unique global optimal solution or if there are multiple local solutions to this optimization problem. 

The optimal solution to the Lagrangian requires the first variation of $\mathcal{L}$ w.r.t. $A$ and $\lambda$ to vanish at these optimal points.

\subsection{First Variation of the Augmented Functional}

Let's derive the first variation of the augmented functional w.r.t. the field variable, $A$.
\[
\begin{aligned}
\delta_A \mathcal{L}(A, \lambda) &= \int_{\Omega} \left[ A\delta A +CA |\nabla A|^2 \delta A+ C|A|^2\nabla A \nabla \delta A + D(\nabla A \nabla \delta A) - F(A) \delta A \right] dV \\
&\quad + \int_{\Omega} \lambda \left( \frac{\partial \delta A}{\partial t} + C (\delta A \cdot \nabla)A + C(A \cdot \nabla)\delta A - D\nabla^2 \delta A - F'(A) \delta A \right) dV
\end{aligned}
\]

Where \( F'(A) \) represents the derivative of the function \(F\) w.r.t. \(A\).

Using appropriate theorems and vector identities, we can write the above expressions as follows and take $\delta A$ common:

\[
\begin{aligned}
\delta_A \mathcal{L}(A, \lambda) &= \int_{\Omega} \left[ A\delta A +CA |\nabla A|^2 \delta A - \left[ \nabla \cdot (C|A|^2 \nabla A) + \nabla \cdot (D \nabla A) \right]  \delta A - F(A) \delta A \right] dV \\
&\quad + \int_{\Omega}  \left (-\delta A \frac{\partial \lambda}{\partial t} - \lambda F'(A) \delta A \right) dV
\end{aligned}
\]

Now, we can write the first variation w.r.t. the Lagrange multiplier, $\lambda$ as follows:

\[
\begin{aligned}
\delta_\lambda \mathcal{L}(A, \lambda) = \int_{\Omega} \delta \lambda \left(\frac{\partial A}{\partial t} + C(A \cdot \nabla)A - D\nabla^2 A - F(A)\right) dV
\end{aligned}
\]
For the optimum solutions, the variations w.r.t. the field variable, $A$ and Lagrange multiplier, $\lambda$ must be zero.

\begin{equation}\label{var_A}
\begin{aligned}
    \delta_A \mathcal{L}(A, \lambda) &= \int_{\Omega} \left[ A\delta A +CA |\nabla A|^2 \delta A - \left[ \nabla \cdot (C|A|^2 \nabla A) + \nabla \cdot (D \nabla A) \right]  \delta A - F(A) \delta A \right] dV \\
&\quad + \int_{\Omega}  \left (-\delta A \frac{\partial \lambda}{\partial t} - \lambda F'(A) \delta A \right) dV = 0
\end{aligned}
\end{equation}

\begin{equation}\label{var_lambda}
\delta_\lambda \mathcal{L}(A, \lambda) = \int_{\Omega} \delta \lambda \left(\frac{\partial A}{\partial t} + C(A \cdot \nabla)A - D\nabla^2 A - F(A)\right) dV = 0
\end{equation}

Equation \eqref{var_A} governs the nature of optimal solution while \eqref{var_lambda} forces the general equation to hold true for any variation in lambda.
For linear or convex energy functional, the solution will be a global minimum. However, for non-convex energy functional, the existence and smoothness of local solutions depend on the magnitude of the convective term compared to the dissipative term. For relatively small convective forces and large diffusive forces, the energy functional will be relatively convex. If $F(A)$ is well defined and does not introduce singularities, the solution in this case will be smooth and potentially global. Intuitively, the stabilizing effect of dissipative forces pushes the initial point to the global optimum. Similar observations can be made for low Reynold's number. \\
On the other hand, for large convective forces, the non-convexity of the energy functional may lead to multiple local solutions. However, the nature of these solutions needs to be studied by solving the above system of equations.

\section{Convexity Analysis and Global Optimum Conditions}
To gain deeper insights into the nature of the solutions of the general equation, let's analyze the structure of energy functional for convexity.

For the positive diffusion coefficient, the first and third terms in the energy functional are quadratic in $A$ and $\nabla A$ respectively. These two terms contribute towards the convexity of energy functional, $E(A)$. The force term may be convex or non-convex depending on the forces associated with the given physical system. In the case of positive convective and diffusion coefficients, and convex $f(A)$ term, the energy functional is likely convex if the interaction between the square of $A$ and $\nabla A$ does not lead to regions of non-convexity. \\
In the cases, when convective and/ or force terms introduce non-convexity and/ or singularity, the optimization framework may lead to the solution which are not global and/ or lack smoothness. For such scenarios, the first variation equation \eqref{var_A} must be further analyzed to gain more understanding of the possible solutions.\\

For an arbitrary value of $\delta A$, the following must be satisfied for $\delta_A \mathcal{L}(A, \lambda)$ to vanish :

\[
\begin{aligned}
  A +CA |\nabla A|^2  - \left[ \nabla \cdot (C|A|^2 \nabla A) + \nabla \cdot (D \nabla A) \right]   - F(A)   -  \frac{\partial \lambda}{\partial t} -\lambda F'(A) =0
\end{aligned}
\]

Rearranging and rewriting the above equation, we get:
\begin{equation}\label{fin_F}
F(A) + \lambda F'(A)  = A +CA |\nabla A|^2  - \left[ \nabla \cdot (C|A|^2 \nabla A) + \nabla\cdot (D \nabla A) \right]   -  \frac{\partial \lambda}{\partial t}
\end{equation}

The external force acting on the system varies with the temporal dependence of $\lambda$ along with convective and diffusive forces in equation \eqref{fin_F} reflects the feedback mechanism embedded in the system as it evolves over time. The presence of a dominant convective term may lead to the erratic behavior of the system. In the presence of such behavior, getting a smooth global optimum for these systems can be challenging.

\section{Incorporating Total Variation Regularization for Smooth Solutions}
To mitigate the noise in the system, we incorporate a smoothness constraint by adding a TV regularization term to our original optimization framework. The resultant TV regularized problem can be solved for global optimality to give a smoother solution compared to the original problem without regularization. 

Let's introduce the TV regularization term in the original energy functional. The modified regularized energy functional, $E_{reg}(A)$ is given by using the regularization parameter, $\lambda_{reg} > 0$ as the following equation:

\begin{equation}
E_{reg}(A) = E(A) + \lambda_{reg} \int_{\Omega}|\nabla A|dV
\end{equation}

 The goal is to minimize the regularized energy functional $E_{\text{reg}}(A)$ subject to the original constraint, \eqref{eq:1} and $\lambda_{reg} > 0$.

Now similar to the non-regularized case, we derive the Euler-Lagrange equation for the regularized energy functional. The first variation of the regularized Lagrangian functional $\mathcal{L}_{\text{reg}}(A, \lambda)$ w.r.t. $A$ and $\lambda$ must be set to zero:

\begin{align}\label{mod_var_alpha_lambda}
\delta_A \mathcal{L}_{\text{reg}}(A, \lambda) &= \int_{\Omega} \left[ A +CA |\nabla A|^2  - \left[ \nabla \cdot (C|A|^2 \nabla A) + \nabla \cdot (D \nabla A) \right]  - F(A)  + \lambda_{reg} \frac{\nabla A}{|\nabla A|}  \right] \delta A  dV + \\
    &\quad + \int_{\Omega}  \left (- \frac{\partial \lambda}{\partial t} - \lambda F'(A)  \right) \delta A dV = 0 \\
\delta_\lambda \mathcal{L}_{\text{reg}}(A, \lambda) &= \int_{\Omega} \left[\frac{\partial A}{\partial t} + C(A \cdot \nabla)A - D\nabla^2 A - F(A) \right] \delta\lambda dV = 0
\end{align}

From the first equation, we obtain the following relationship:

\begin{equation}
F(A) + \lambda F'(A) = A +CA |\nabla A|^2  - \left[ \nabla \cdot (C|A|^2 \nabla A) + \nabla\cdot (D \nabla A) \right]   -  \frac{\partial \lambda}{\partial t} + \lambda_{reg} \frac{\nabla A}{|\nabla A|}
\end{equation}

The regularization term improves the smoothness of the solution. This can reduce the noise in the feedback mechanism embedded in the system. The reduced noise causes the system to stabilize faster leading to relatively smoother solutions compared to the non-regularized solutions. The regularized energy function can be specifically utilized to gain insights into the solution when convective forces are dominant.

Taking inspiration from Ricci flow, we propose a gradient flow similar to Ricci flow to update variable, $A$, but for optimization problems. We develop an iterative process that updates the variable, $A$ towards minimizing the modified energy functional, $E_{\text{reg}}(A)$.  We define the evolution of A in terms of the derivative of regularized energy functional w.r.t. $A$.

\begin{equation}
\frac{\partial A}{\partial t} = -\frac{\delta E_{\text{reg}}(A)}{\delta A}
\end{equation}

Now, the derivative of $E_{\text{reg}}(A)$ w.r.t. $A$ can be obtained from the equation \eqref{mod_var_alpha_lambda} by isolating the terms involving $\delta A$. We have:

\begin{equation}
\frac{\delta E_{\text{reg}}(A)}{\delta A} = \int_{\Omega} \left[ A +CA |\nabla A|^2  - \left[ \nabla \cdot (C|A|^2 \nabla A) + \nabla \cdot (D \nabla A) \right]  - F(A)  + \lambda_{reg} \frac{\nabla A}{|\nabla A|} \right] dV
\end{equation}

Now, the value of field variable, A at $k+1^{th}$ iteration towards minimizing energy functional, $E_{\text{reg}}(A)$ can be written as the following equation:

\begin{equation}\label{iter}
A_{k+1} = A_{k} - \tau_k \left[ A_k +CA_k |\nabla A_k|^2  - \left[ \nabla \cdot (C|A_k|^2 \nabla A_k) + \nabla \cdot (D \nabla A_k) \right]  - F(A_k)  + \lambda_{reg} \frac{\nabla A_k}{|\nabla A_k|}\right],
\end{equation}

where $\tau_k > 0$ is a time step parameter.

This iterative process updates the field variable $A$ towards minimizing the regularized energy functional $E_{\text{reg}}(A)$. This flow is conceptually similar to Ricci Flow, albeit, in a different mathematical setting, as it evolves the variable $A$ towards minimizing the regularized energy functional.

\section{Alternative Bounds for the Solution Space}

Let's rewrite the iterative process derived in the equation \eqref{iter} in the following form:

\begin{equation}
A_{k+1} = A_{k} - \tau_k (P_k + \lambda_{reg} Q_k),
\end{equation}

where the equations for $P_k$ and $Q_k$ are:

\begin{align}
P_k &= A_k + CA_k |\nabla A_k|^2 - \nabla \cdot (C|A_k|^2 \nabla A_k) - \nabla \cdot (D \nabla A_k) -F(A_k), \\
Q_k &=  \frac{\nabla A_k}{|\nabla A_k|}
\end{align}

Now, let's assume the following:
\begin{equation}\label{assum}
    \nabla A \geq 0
\end{equation}

Given the assumption in equation \eqref{assum}, we can consider the following to be the lower bound of $P_k$:

\begin{align}
P_k^{L} &= A_k + CA_k |\nabla A_k|^2 - \nabla \cdot (C|A_k|^2 \nabla A_k) - \nabla \cdot (D \nabla A_k) -  \frac{\nabla A_k}{|\nabla A_k|}
\end{align}

Similarly, for the upper bound of $P_k$, we may write the following:

\begin{align}
P_k^{U} &= A_k + CA_k |\nabla A_k|^2 - \nabla \cdot (C|A_k|^2 \nabla A_k) - \nabla \cdot (D \nabla A_k) +  \frac{\nabla A_k}{|\nabla A_k|}
\end{align}

We may consider the following lower and upper bounds of $Q_k$:

\begin{align}
Q_k^{L} &= 0,
\end{align}
\begin{align}
Q_k^{U} &= 2\frac{\nabla A_k}{|\nabla A_k|}
\end{align}

Now, we may write the following system of inequalities for the convex lower and upper bounds of $P_k$ and $Q_k$:

\begin{align}
P_k^{L} &\leq P_k \leq P_k^{U}, \\
Q_k^{L} &\leq Q_k \leq Q_k^{U}.
\end{align}

We can derive the corresponding updated equations for the lower and upper bounds of $A_{k+1}$ using the alternative convex lower and upper bounds. For the lower bound, we have:

\begin{equation}
A_{k+1}^{L} = A_{k} - \tau_k (P_k^{L} + \lambda_{reg} Q_k^{L}).
\end{equation}

Similarly, for the upper bound, we have:

\begin{equation}
A_{k+1}^{U} = A_{k} - \tau_k (P_k^{U} + \lambda_{reg} Q_k^{U}).
\end{equation}

Let's now assess the tightness of bounds as it constitutes the solution space:

\begin{align}
\Delta A_{k+1} = A_{k+1}^{U} - A_{k+1}^{L} = \tau_k ((P_k^{U} - P_k^{L}) + \lambda_{reg} (Q_k^{U} - Q_k^{L})).
\end{align}

Let's substitute the expressions for $P_k^{L}, P_k^{U}, Q_k^{L},$ and $Q_k^{U}$ into the equation for $\Delta A_{k+1}$:

\begin{align}
\Delta A_{k+1} &= -\tau_k ((P_k^{U} - P_k^{L}) + \lambda_{reg} (Q_k^{U} - Q_k^{L})) \\
&= -2\tau_k (1+\lambda_{reg} ) \frac{\nabla A_k}{|\nabla A_k|}.
\end{align}

The tight solution space for small values of the time step, $\tau_k$, and regularization parameter, $\lambda_{reg}$, provides a way to approximate the solution of the optimization problem. Furthermore, these bounds provide a good starting point for finding the global optimum. 

Please note that the above analysis can be adapted for the case when our initial assumption (equation \eqref{assum}) does not hold, i.e., when $\nabla A <0 $. In that case, the upper and lower bounds for each $P_k$ and $Q_k$ are interchanged leading to the same mathematical expression for $\nabla A_{k+1}$ except for a positive sign.

\section{Convergence Analysis}
Let's assume that the energy function, $E_{reg}(A)$ is convex w.r.t A and there exists a unique global minimum for the regularized optimization framework. 

As we update the variable, $A$ using the iterative equation \eqref{iter} the energy functional changes for some Lipschitz constant, $L > 0$, as per the descent lemma \cite{huang2022cubic} as the following equation: 

\begin{align}
E_{\text{reg}}(A_{k+1}) - E_{\text{reg}}(A_k) &= \left[ E_{\text{reg}}(A_k - \tau_k(P_k + \lambda_{reg} Q_k)) \right] - E_{\text{reg}}(A_k) \\
&\leq -\tau_k \left( \frac{\delta E_{\text{reg}}(A_k)}{\delta A_k} \right)^T (P_k + \lambda_{reg} Q_k) + \frac{L\tau_k^2}{2} ||P_k + \lambda_{reg} Q_k||^2, \forall P_k, Q_k
\end{align}

Given the convexity of the energy functional $E_{\text{reg}}(A)$ as we update the variable, $A$, with every iteration, energy functional decreases, i.e., 

\begin{equation}
E_{\text{reg}}(A_{k+1}) \leq E_{\text{reg}}(A_k).
\end{equation}

Now, with every iteration, the gap between successive iterations changes as follows:

\begin{align}
\Delta A_{k+1} - \Delta A_{k} 
&= -2(1+\lambda_{reg}) (\tau_{k+1} \frac{\nabla A_{k+1}}{|\nabla A_{k+1}|}-\tau_k \frac{\nabla A_k}{|\nabla A_k|}).
\end{align}

Since the energy functional decreases over iterations,  the magnitude of the gap between successive iterations must decrease to zero over iterations for the solution to converge to the global optimum:
\begin{equation}
|\Delta A_{k+1} - \Delta A_k| \leq 0.
\end{equation}

As we iterate from one step to the next, the gap between successive iterations must decrease for the solution to converge to the global optimum. As this gap decreases, the solution must converge to the global optimum given the convexity of the energy functional and the existence of a unique global optimum.

\section{Approximate Solutions for Physical Systems}

Now, we will demonstrate the application of our method on well-known physical systems.

\subsection{Boltzmann equation}

In statistical mechanics, the evolution of particle distribution function, $f(x, \xi, t)$ at location, x, velocity $\xi$ and time, $t$ can be described using Boltzmann transport equation \cite{mohamad2011boltzmann} as follows:

\begin{equation}
\frac{\partial f}{\partial t} + \mathbf{\xi} \cdot \nabla f + \frac{\mathbf{F}}{\rho} \cdot \nabla_{\mathbf{\xi}} f = \Omega (f),
\end{equation}

where \( \mathbf{F} \) is an external force, \( \rho \) is the particle density, and \( \Omega(f) \) represents the collision operator.

The Energy functional for the Boltzmann equation can be written in terms of kinetic energy, entropy, and a function governing external forces and collision operator as follows:
\[
E(f) = \int \int \left[ \frac{1}{2} m |\xi|^2 f(x, \xi, t) - k_B T f(x, \xi, t) \log(f(x, \xi, t)) + \Psi(x, \xi, f) \right] d\xi \, dx,
\]

where \(\Psi(x, \xi, f)\) is a function of external forces and collision operator. The structure of the function can be written for the specific problem at hand.

The gradient flow for this problem can be written as follows:

\[
\frac{\partial f}{\partial t} = -\frac{\delta E_{\text{reg}}(f)}{\delta f}
 = \frac{1}{2} m |\xi|^2 - k_B T (1 + \log(f)) + \frac{\delta \Psi}{\delta f}
\]

The TV regularized energy functional term for the above equation, analogous to that for our generalized equation, can be written as follows:

\begin{equation}
E_{\text{reg}}(f) = E(f) + \lambda_{reg} \int |\nabla f| d\mathbf{r} d\mathbf{v},
\end{equation}

The iterative update for minimizing \( E_{\text{reg}}(f) \) is:

\begin{equation}
f_{k+1} = f_{k} - \tau_k \left[ ( \frac{1}{2} m |\xi|^2 - k_B T (1 + \log(f)) + \frac{\delta \Psi}{\delta f} + \lambda_{reg} \nabla \cdot \left( \frac{\nabla f_{k}}{|\nabla f_{k}|} \right) \right],
\end{equation}

where \( \tau_k \) is a step size parameter. 

In the steady-state solution where \( \frac{\partial f}{\partial t} = 0 \), the Boltzmann equation becomes:

\begin{equation}
0 =  \frac{1}{2} m |\xi|^2 - k_B T (1 + \log(f)) + \frac{\delta \Psi}{\delta f} - \lambda_{reg} \nabla \cdot \left( \frac{\nabla f}{|\nabla f|} \right).
\end{equation}

The steady-state approximate can be further simplified using the BGK operator \cite{bhatnagar1954model} given the structure of function $\psi$.

\subsection{Compressible Navier-Stokes Equation}

The behavior of compressible fluid with density, $\rho$, velocity, $v$ under pressure $p$ and stress tensor, $S$ in fluid dynamics can be modeled using the Navier-Stokes equations \cite{lukaszewicz2016navier} as follows:

\begin{equation}
\frac{\partial \rho}{\partial t} + \nabla \cdot (\rho \mathbf{v}) = 0,
\end{equation}

\begin{equation}
\frac{\partial (\rho \mathbf{v})}{\partial t} + \nabla \cdot (\rho \mathbf{v} \otimes \mathbf{v} + p\mathbf{I}) = \nabla \cdot \mathbf{S},
\end{equation}

where  \(\mathbf{I}\) is the identity matrix.

We can write the energy functional as follows:

\[
E(\mathbf{v}, \rho, p) = \frac{1}{2} \int_{\Omega} (\rho |\mathbf{v}|^2  + C \int_{\Omega} \rho |\mathbf{v}|^2 |\nabla \mathbf{v}|^2  + D  |\nabla \mathbf{v}|^2) dV  - \int_{\Omega} f(\rho, p)) dV,
\]

The regularized energy functional \(E_{\text{reg}}(\rho, \mathbf{v})\) can be written using regularization parameters for density and velocity i.e., $\lambda_{reg, \rho}$ and $\lambda_{reg, v}$:

\begin{equation}
E_{\text{reg}}(\rho, \mathbf{v}) = E(\rho, \mathbf{v}) + \lambda_{reg, \rho} \int |\nabla \rho|dV + \lambda_{reg, v} \int |\nabla \mathbf{v}|dV,
\end{equation}

The iterative process for minimization:

\begin{equation}
\rho_{k+1} = \rho_{k} - \tau_k \left[ -\partial E_{reg}/ \partial \rho + \lambda_{reg, \rho} \nabla \cdot \left( \frac{\nabla \rho_{k}}{|\nabla \rho_{k}|} \right) \right],
\end{equation}

\begin{equation}
\mathbf{v}_{k+1} = \mathbf{v}_{k} - \tau_k \left[ -\partial E_{reg}/ \partial \mathbf{v} + \lambda_{reg, \mathbf{v}} \nabla \cdot \left( \frac{\nabla \mathbf{v}_{k}}{|\nabla \mathbf{v}_{k}|} \right) \right],
\end{equation}

We can apply steady state approximation and perturbation analysis to gain further insights into the smoothness and existence of the solution. 

\subsection{Maxwell Equation}
In elcetromagnestism, the behaviour of electric (\(\mathbf{\phi_E}\)) and magnetic (\(\mathbf{\phi_B}\)) fields due to charge density, $\rho$ and current density, $J$ is governed by the following four Maxwell equations \cite{haus1989electromagnetic}:

\textbf{Gauss's Law} (electricity):
\begin{equation}
\nabla \cdot \mathbf{\phi_E} = \frac{\rho}{\epsilon_0},
\end{equation}

\textbf{Gauss's Law} (magnetism):
\begin{equation}
\nabla \cdot \mathbf{\phi_B} = 0,
\end{equation}

\textbf{Faraday's Law}:
\begin{equation}
\nabla \times \mathbf{\phi_E} = -\frac{\partial \mathbf{\phi_B}}{\partial t},
\end{equation}

\textbf{Ampere's Law} (with Maxwell's addition):
\begin{equation}
\nabla \times \mathbf{\phi_B} = \mu_0 \mathbf{J} + \mu_0 \epsilon_0 \frac{\partial \mathbf{\phi_E}}{\partial t},
\end{equation}

where \(\epsilon_0\) and \(\mu_0\) are vacuum permittivity and vacuum permeability respectively.

The energy functional for Maxwell's equation can be defined as follows:

\[
E(\mathbf{\phi_E}, \mathbf{\phi_B}) = \frac {1}{2}\int_V (\epsilon_0 |\phi_E| ^2 +\frac{1}{\mu_0} |\phi_B|^2)dV
\]

The regularized energy functional \(E_{\text{reg}}(\mathbf{\phi_E}, \mathbf{\phi_B})\) is

\begin{equation}
E_{\text{reg}}(\mathbf{\phi_E}, \mathbf{\phi_B})) = E(\mathbf{\phi_E}, \mathbf{\phi_B})) + \lambda_{{reg}, \phi_E} \int |\nabla \mathbf{\phi_E}|  dV + \lambda_{{reg}, \phi_B} \int |\nabla \mathbf{\phi_B}|dV,
\end{equation}

The iterative process to minimize \(E_{\text{reg}}\) is:

\begin{equation}
\mathbf{\phi_E}_{k+1} = \mathbf{\phi_E}_k - \tau_{k,E} \left[ -\frac{\partial E_{reg}}{\partial {\phi_E}} + \lambda_{{reg}, \phi_E} \nabla \cdot \left( \frac{\nabla \mathbf{\phi_E}_k}{|\nabla \mathbf{\phi_E}_k|} \right) \right],
\end{equation}

\begin{equation}
\mathbf{\phi_B}_{k+1} = \mathbf{\phi_B}_k - \tau_{k,B} \left[ -\frac{\partial E_{reg}}{\partial {\phi_B}} + \lambda_{{reg}, \phi_B} \nabla \cdot \left( \frac{\nabla \mathbf{\phi_B}_k}{|\nabla \mathbf{\phi_B}_k|} \right) \right],
\end{equation}

The above iterative process allows for minimizing energy function, $E_{reg}$ by optimizing the field variables while maintaining the smoothness in the fields.

\section{Conclusion}
We proposed an optimization framework around a generalized equation governing physical systems such as fluid dynamics, statistical mechanics, and other related physical systems. We introduced an energy functional for the generalized equation. The objective of the optimization problem is to minimize the energy functional by optimizing the primary variable. The objective function (the energy functional) and the constraint (the generalized equation) are written as a linear combination using a Lagrange multiplier. The resultant Euler-Lagrange equation is further investigated using the first variation for convexity and global optimum. \\
Furthermore, the proposed method is refined for smooth solutions by incorporating the TV regularization term for the primary variable in the original energy functional. Using this modified energy functional, we develop a flow similar to the Ricci flow, aimed towards minimizing the modified energy function by optimizing the primary variable. The optimal value of the primary variable is obtained through an iterative process by using the gradient descent method.\\
The mathematical structure of the iterative equation is used to develop lower and upper bounds on solution space. These bounds can provide a good starting point for the heuristic approach for finding a solution for the optimization. A thorough convergence analysis is carried out to assess the utility of the approach. \\
Finally, we illustrate the application of this method by applying it to statistical mechanics, fluid dynamics, and electromagnetism. 

\section{Declarations}

\subsection{Ethics approval and consent to participate}
Not applicable. This study did not involve human participants, human data, or human tissue.

\subsection{Consent for publication}
Not applicable. This manuscript does not contain data from any individual person.

\subsection{Availability of data and materials}
Not applicable. No new data were created or analyzed in this study.

\subsection{Competing interests}
The authors declare that they have no competing interests.

\subsection{Funding}
This research did not receive any specific grant from funding agencies in the public, commercial, or not-for-profit sectors.

\subsection{Authors' contributions}
VG contributed entirely to the conception and design of the study, performed the analysis, and drafted the manuscript. All authors read and approved the final manuscript.

\subsection{Acknowledgements}
Not applicable.

\newpage
\bibliography{aomsample}
\bibliographystyle{unsrt}

\end{document}